\documentclass[fleqn,usenatbib]{mnras}

\usepackage[T1]{fontenc}
\usepackage{ae,aecompl}

\usepackage{graphicx}	
\usepackage{amsmath}	
\usepackage{amssymb}	
\usepackage{pdflscape}
\usepackage{times,txfonts}
\usepackage{color}
\usepackage{hyperref}	

\definecolor{linkcolor}{rgb}{0,0,0.25}
\hypersetup{
  colorlinks=true,        
  linkcolor=linkcolor,    
  citecolor=linkcolor,    
  filecolor=linkcolor,    
  urlcolor=linkcolor      
}

\usepackage{multirow}
\usepackage{rotating}
\usepackage{lscape}
\usepackage{tablefootnote}
\usepackage{hyperref}
\usepackage{listings}

\usepackage[colorinlistoftodos]{todonotes}

\definecolor{salmon}{rgb}{0.95,0.5,0.25}

\newcommand{\dalpha}{$\rm d \alpha / d \ln (|D_\mathrm{prog}|)$}
\title[The Stellar Mass Function Along Stellar Streams]{Variation in the Stellar Mass Function Along Stellar Streams}
\author[Webb \& Bovy]{Jeremy J. Webb\thanks{E-mail: webb@astro.utoronto.ca (JW)} \& Jo Bovy
 \\
David A. Dunlap Department of Astronomy and Astrophysics, University of Toronto, 50 St. George Street, Toronto, ON, M5S 3H4, Canada
}

\begin{document}

\pagerange{\pageref{firstpage}--\pageref{lastpage}} \pubyear{2021}

\maketitle

\label{firstpage}

\begin{abstract}
Stellar streams are the inevitable end product of star cluster evolution, with the properties of a given stream being related to its progenitor. We consider how the dynamical history of a progenitor cluster, as traced by the evolution of its stellar mass function, is reflected in the resultant stream. We generate model streams by evolving star clusters with a range of initial half-mass relaxation times and dissolution times via direct $N$-body simulations. Stellar streams that dissolve quickly show no variation in the stellar mass function along the stream. Variation is, however, observed along streams with progenitor clusters that dissolve after several relaxation times. The mass function at the edges of a stream is approximately primordial as it is populated by the first stars to escape the cluster before segregation occurs. Moving inwards the mass function steepens as the intermediate parts of the stream consist of mostly low-mass stars that escaped the cluster after some segregation has occurred. The centre of the stream is then marked by a flatter mass function, as the region is dominated by high-mass stars that quickly segregated to the progenitor cluster's centre and were the last stars to become unbound. We further find that the maximum slope of the mass function along the stream and the rate at which it decreases with distance from the dissolved progenitor serve as proxies for the dynamical state reached by the progenitor cluster before dissolution; this may be able to be applied to observed streams with near-future observations.

\end{abstract}


\section{Introduction} \label{s_intro}

Stellar streams represent the final stage of star cluster evolution, as stellar evolution, two-body interactions, and tidal stripping all lead to cluster dissolution \citep[e.g.,][]{Baumgardt03,Heggie03}. The fact that star clusters evolve within the external tidal field of their host galaxies causes some cluster stars to become energetically unbound from the cluster and escape via one of the first two Lagrange points of the cluster--galaxy system \citep{Binney08, Renaud2011}. Stars that escape do so with low velocity, which over time results in leading and trailing cold tidal tails emanating from a cluster's tidal radius. A large number of cold stellar streams have been observed in the Milky Way \citep{grillmair16, Mateu2017}, with data releases from \emph{Gaia} \citep{Gaia16a,gaia18} aiding in the ability to detect stellar streams with the help of stellar kinematics \citep[e.g.][]{Malhan2018, Ibata2019, Starkman2020, Borsato2020, Ibata2021}. It is notable that many of the known cold stellar streams in the Milky Way do not have an obvious progenitor cluster (e.g., GD-1; \citealt{grillmair06}). 

Given that stars escape a star cluster through its Lagrange points, key properties of a stream's progenitor are encoded within the stream's properties. For example, the total mass and velocity dispersion of a stream's progenitor can be inferred from its cross-section and luminosity as stars escaping high-mass progenitors do so with a higher velocity dispersion, resulting in thicker and longer streams \citep{Johnston1998, Gialluca2021}. A progenitor's disruption time also affects the stream's length, width, and density as longer disruption times allow for stream stars to travel larger distances along and perpendicular to the stream, resulting in a continual decrease in stream density over time \citep{Helmi1999}. These behaviours have allowed for constraints to be placed on the mass and size of several stream progenitors with the help of $N$-body simulations of star cluster disruption \citep[e.g.,][]{Sales2008, Webb2019}. However, these approaches are limited by the necessary assumption that the stream represents the entire progenitor cluster and that the progenitor cluster's orbit in the Milky Way is constant with time.

Density variations have also been observed along stellar streams, which may occur naturally due to the epicyclic motion of stars that escape the progenitor cluster \citep{Kupper2008, Just2009, kupper11, Kupper2015, Ibata2020} Such variations could then, in principle, be used to constrain the progenitor's properties. However, interactions between streams and various forms of baryonic and dark matter substructure \citep[e.g.,][] {Ibata2001, johnston02, erkal15a, erkal15b, Amorisco16a, bovy17, pearson17, banik18, Bonaca2019} also lead to fluctuations in stellar density along stellar streams, as would periodic mass loss due to the cluster having an eccentric or inclined orbit. Furthermore, an underdensity is also expected along the stream at the location of the now dissolved progenitor if dissolution occurred recently \citep{PriceWhelan2018, Webb2019}. Hence there is a large amount of degeneracy in using density fluctuations along stellar streams to robustly constrain the stream's progenitor.

One property of stellar streams that has yet to be explored, primarily due to observational limitations, is the distribution of stellar masses along the stream. As star clusters evolve, two-body interactions between members result in star clusters evolving towards a state of partial energy equipartition \citep{Merritt1981, Miocchi2006, Trenti2013, Gieles2015, Bianchini2016}. A bi-product of this evolution is that star clusters become mass segregated, as high-mass stars fall towards a cluster's centre and low-mass stars migrate outwards \citep{Heggie03}. Hence, in the absence of primordial mass segregation \citep{Baumgardt2008, Vesperini2009},  dynamically young clusters will have undergone little to no mass segregation while dynamically old clusters will have high degrees of mass segregation. 

Through a suite of direct $N$-body simulations, \citet{Webb2016} illustrated that measuring the radial variation in the slope $\alpha$ of a cluster's stellar mass function (MF) provides a strong indication of the cluster's dynamical state. This result was confirmed by \citet{Baumgardt2017} by comparing $N$-body simulations to Galactic globular clusters. \citet{Webb2016} find that while initially the slope of the MF shows no radial gradient ($\alpha(r) \sim$ constant, where $r$ is distance from the cluster centre), $\alpha(r)$ eventually starts to increase in the central regions as high-mass stars fall inwards and decrease in the outer regions as low-mass stars migrate outwards. The slope of this radial gradient---measured as $\delta \alpha = \mathrm{d} \alpha(r) /  \mathrm{d} \ln(r/r_m) $ where $r_m$ is the cluster's half-mass radius---decreases as a function of time at a rate that scales with the cluster's half-mass relaxation time $t_{rh}$. 

Eventually, as the outer regions of a cluster that are stripped by tides become primarily populated with low-mass stars over several relaxation times, the preferential escape of low-mass stars causes a cluster's global mass function to change \citep{Vesperini1997, Kruijssen2009, Gieles2011, Webb2015, Balbinot2018}. Once a cluster's global $\alpha$ starts increasing, \citet{Webb2016} further find that the rate at which $\delta \alpha$ decreases will slow until it reaches a constant value due to the outwards migration of low-mass stars being balanced by the rate at which low-mass stars are being stripped from the cluster.

The time evolution of $\alpha(r)$ within a cluster will directly influence the mass function of escaping stars as well. Since stars primarily escape from the outer regions of a cluster where we expect $\alpha$ to decrease with time, we expect the mass function to vary along a stellar stream. The amount of variation should depend on the rate at which the progenitor cluster was able to segregate and the rate at which it loses mass. Similarly once the cluster has reached dissolution, the $\delta \alpha$ achieved by the progenitor cluster will also leave an imprint via variation in the mass function near the centre of the stream.  

In this study, through a suite of direct $N$-body simulations of star cluster evolution, we determine how the distribution of stellar masses along a stellar stream is related to the dynamical age of the stream's progenitor. Establishing a relationship between variation in the mass function along a stellar stream and its progenitor's dynamical age will provide estimates of a stream progenitor's properties at formation that don't require assumptions to be made regarding the progenitor's orbital history or that the entire stream be recovered. Such estimates will be complimentary to methods that make use of the stream's length, width, velocity dispersion and density that do require these additional assumptions \citep[e.g.][]{Balbinot2018,Webb2019, Gialluca2021}. In Section \ref{s_nbody}, we introduce and explain the suite of simulations. In Section \ref{s_results}, we explore how the mean mass and stellar mass function changes as both a function of stellar escape times and a star's location along the stream after the progenitor cluster has dissolved. We further explore in Section \ref{s_discussion} how our results can be used to constrain the progenitors of observed stellar streams and summarise our findings in Section \ref{s_conclusion}.

\section{Simulations} \label{s_nbody}

In order to test how the degree of mass segregation reached by a given cluster is reflected in the distribution of stellar masses along a stream, we perform a suite of direct $N$-body simulations of star cluster evolution. Direct $N$-body simulations are required in order to accurately model two-body interactions and a large suite is necessary in order to investigate a wide range of initial relaxation times and initial dissolution times. We therefore make use of the direct $N$-body code \texttt{NBODY6} \citep{aarseth03}, where we evolve clusters with a range of initial masses, sizes, and orbital distances to dissolution. 

The initial positions and velocities of stars in each of our model clusters are drawn from a Plummer model \citep{Plummer1911}. Stellar masses are sampled between 0.1 and 50 $M_{\odot}$ from a \citet{Kroupa93} initial MF, which has the form:

\begin{equation}
    \frac{dN}{dm} = m^{\alpha}
\end{equation}

were $\alpha$ is equal to $-2.7$ for m > 1 $M_{\odot}$, $-2.2$ for 0.5 $M_{\odot}$ < m <
1 $M_{\odot}$, and $-1.3$ for 0.08 < m < 0.5 $M_{\odot}$. The stellar evolution of single stars follows the prescriptions of \citet{Hurley2000} for a metallicity of Z=0.001, with any binaries that form in the simulation following the prescriptions of \citet{Hurley2002}. The initial binary fraction, however, is assumed to be zero. For the external tidal field in which all clusters orbit we use the Milky-Way model \texttt{MWPotential2014} from \citet{bovy15}, which consists of a spherical power-law bulge with exponential cut-off, stellar disk \citep{miyamoto75}, and dark-matter halo \citep{Navarro96}. 

In order to fully investigate how well a cluster's dynamical state is imprinted on its stellar stream, clusters with a wide range of initial relaxation times and dissolution times must be simulated. We therefore consider clusters that initially have either 12,500 or 25,000 stars, initial half-mass radii or 2 pc, 4 pc, 6 pc or 8 pc, and that have circular orbits at distances of 5 kpc or 10 kpc in the plane of the disk. Each cluster is evolved until the total number of energetically bound stars is less than 100, which we consider to be dissolution. All models are listed in Table \ref{table:models}, along with their dissolution times $t_\mathrm{diss}$ and initial relaxation times $t_{rh,0}$. We adopt a naming convention that reflects the initial number of stars in a cluster, its initial size, and orbital distance. For example, a cluster within initially 12,500 stars, a circular orbit distance of 5 kpc, and an initial half-mass radius of 4 pc, and  is named n12rgc5rm4.

\begin{table}
\centering
\begin{tabular}{l|ccccc}
Model Name & $N_0$ & $r_{gc}$ & $r_{h,0}$ & $t_{rh,0}$ & $t_\mathrm{diss}$  \\
\hline
{} & {} & {$\rm kpc$} & {$\rm pc$} & {$\rm Myr$} & {$\rm Myr$}\\
\hline

n12rgc5rh2 & 12500 & 5 & 2 & 107.9 & 3720.0 \\
n12rgc5rh4 & 12500 & 5 & 4 & 291.6 & 2620.0 \\
n12rgc5rh6 & 12500 & 5 & 6 & 549.6 & 540.0 \\
n12rgc5rh8 & 12500 & 5 & 8 & 848.9 & 300.0 \\
n12rgc10rh2 & 12500 & 10 & 2 & 103.7 & 7280.0 \\
n12rgc10rh4 & 12500 & 10 & 4 & 295.2 & 7500.0 \\
n12rgc10rh6 & 12500 & 10 & 6 & 541.3 & 4940.0 \\
n12rgc10rh8 & 12500 & 10 & 8 & 835.7 & 1980.0 \\
n25rgc5rh2 & 25000 & 5 & 2 & 136.8 & 5980.0 \\
n25rgc5rh4 & 25000 & 5 & 4 & 388.1 & 5480.0 \\
n25rgc5rh6 & 25000 & 5 & 6 & 719.1 & 1740.0 \\
n25rgc5rh8 & 25000 & 5 & 8 & 1109.2 & 500.0 \\
n25rgc10rh4 & 25000 & 10 & 4 & 391.2 & 12160.0 \\
n25rgc10rh6 & 25000 & 10 & 6 & 719.1 & 11240.0 \\
n25rgc10rh8 & 25000 & 10 & 8 & 1101.2 & 7640.0 \\
\hline
\end{tabular}
\caption{Summary table of direct $N$-body simulations used in this study, listing model name, initial number of stars, circular orbit distance, effective radius, initial relaxation time and dissolution time.}
\label{table:models}
\end{table}

\begin{figure}
    \includegraphics[width=0.48\textwidth]{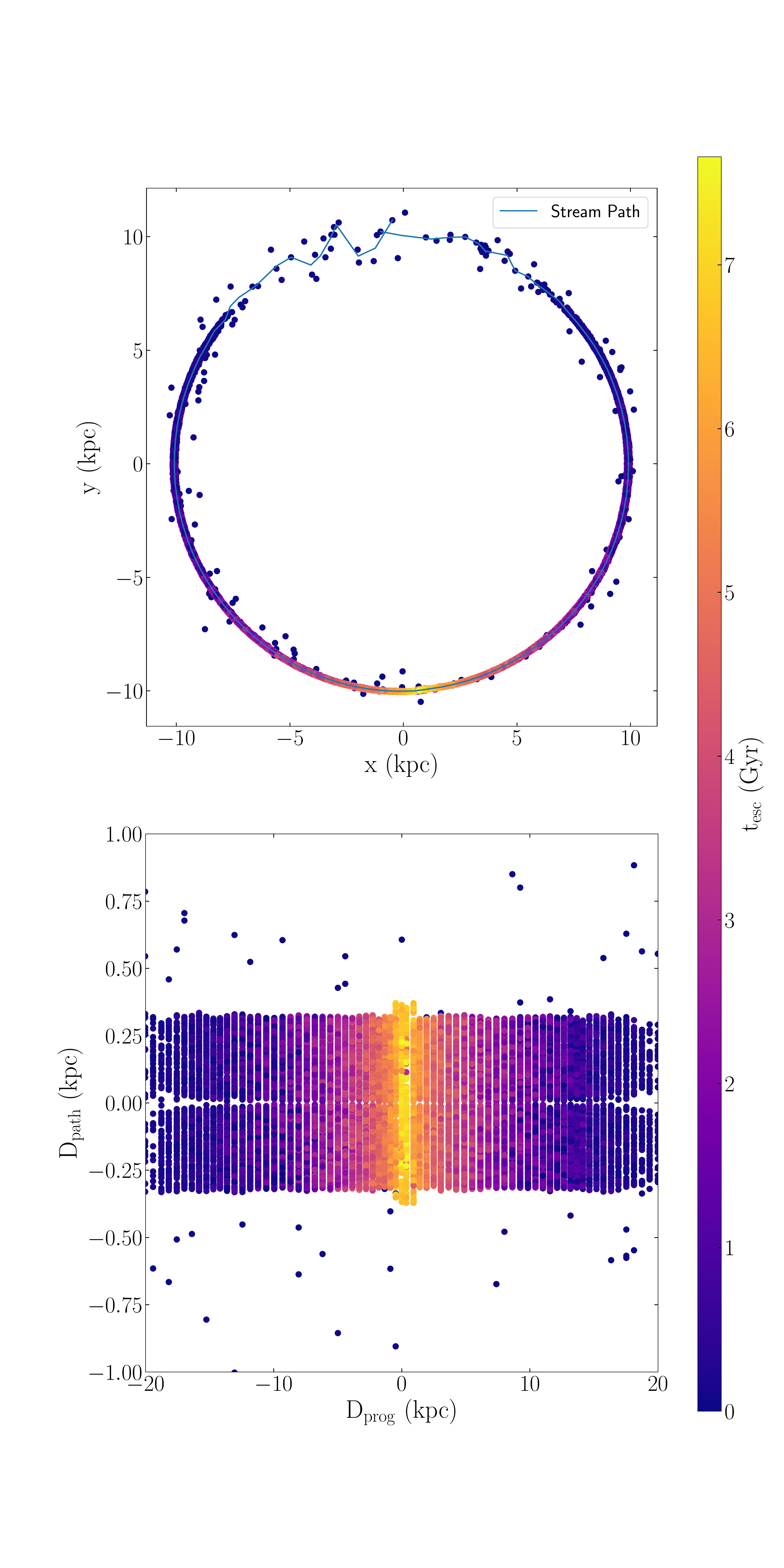}
    \caption{Top Panel: $x$ and $y$ position of stream stars, in galacotcentric coordinates, for model n25rgc10rm8. Stars are colour coded based on the time at which they escape the cluster, as indicated by the colour bar, and the stream path is shown in blue. Lower Panel: Distance from stream path as a function of distance along the stream path from the stream centre, where the now the dissolved progenitor would be located. Stars closest to the stream centre escaped at times comparable to the cluster's dissolution time ($\sim 8$ Gyr) while stars found at the edges of the stream escaped near the beginning of the simulation ($t \sim 0$ Gyr).}
    \label{fig:n25rgc10rm8}
\end{figure}

Over the course of each model cluster's dissolution, each star's escape time is recorded to be the final timestep at which it becomes energetically unbound, as some stars become energetically unbound and are then re-captured. Once a cluster reaches dissolution, we construct a mean stream path by first finding the closest point along the progenitor's orbital path to each star. Stars are then binned based on the point along the orbital path that they are closest to and we determine the mean position in each bin. As illustrated in the top panel of Figure~\ref{fig:n25rgc10rm8}, the mean positions in each bin correspond to a mean stream path. Once the mean stream path is constructed, the distance of each stars from the path $D_{path}$ is determined with stars below the plane of the disk assigned negative values. The distance of each star from the would-be location of the progenitor, as measure along the tail path, is taken to be $D_\mathrm{prog}$. As illustrated in the lower panel of Figure \ref{fig:n25rgc10rm8} for model n25rgc5rm8, this representation allows all streams to be compared within a similar parameter space. Note that the vertical striping in $D_{prog}$ and apparent horizontal gap at $D_{path}$ in the lower panel of Figure \ref{fig:n25rgc10rm8} are both artifacts of how finely we sample the orbital path when constructing the stream path. Also illustrated in Figure \ref{fig:n25rgc10rm8} is the escape time of each star, which confirms the obvious interpretation that stars close to the progenitor's location escaped recently ($t \sim 8$ Gyr, the cluster's dissolution time), while stars near the edges of the stream escaped long ago ($t \sim 0$ Gyr) .

\section{Results} \label{s_results}

In order to study how a progenitor cluster's dynamical state is reflected in the properties of its stellar stream, we first consider the evolution of the mean mass and stellar mass function of escaping stars as a function of escape time in Section \ref{ss_tesc}. In Section \ref{ss_dprog}, we determine how escape time maps onto a star's location on a stellar stream to determine if variations in mean mass and the stellar mass function along the stream can be used to constrain the dynamical age of the stream's progenitor. In each of these subsections, we primarily use the models that initially consist of 25,000 stars to illustrate trends between stream properties and both the cluster's dissolution time and initial relaxation time, because they nicely span the $t_\mathrm{diss}/t_{rh,0}$ parameter space, which we use as a proxy for the progenitor's dynamical state at dissolution. Other commonly used proxies, such as $t_\mathrm{diss}/t_{rh}(t_{diss})$ or $t_\mathrm{diss}/\int_0^{t_{diss}} t_{rh}(t)$ are not applicable when the cluster is at dissolution. Including models with initially 12,500 stars in each figure makes it difficult to observe trends with $t_\mathrm{diss}/t_{rh,0}$, however all models are included when comparing the entire suite of simulations. 

\subsection{Properties of Stars as a Function of Escape Time} \label{ss_tesc}

\begin{figure}
    \includegraphics[width=0.48\textwidth]{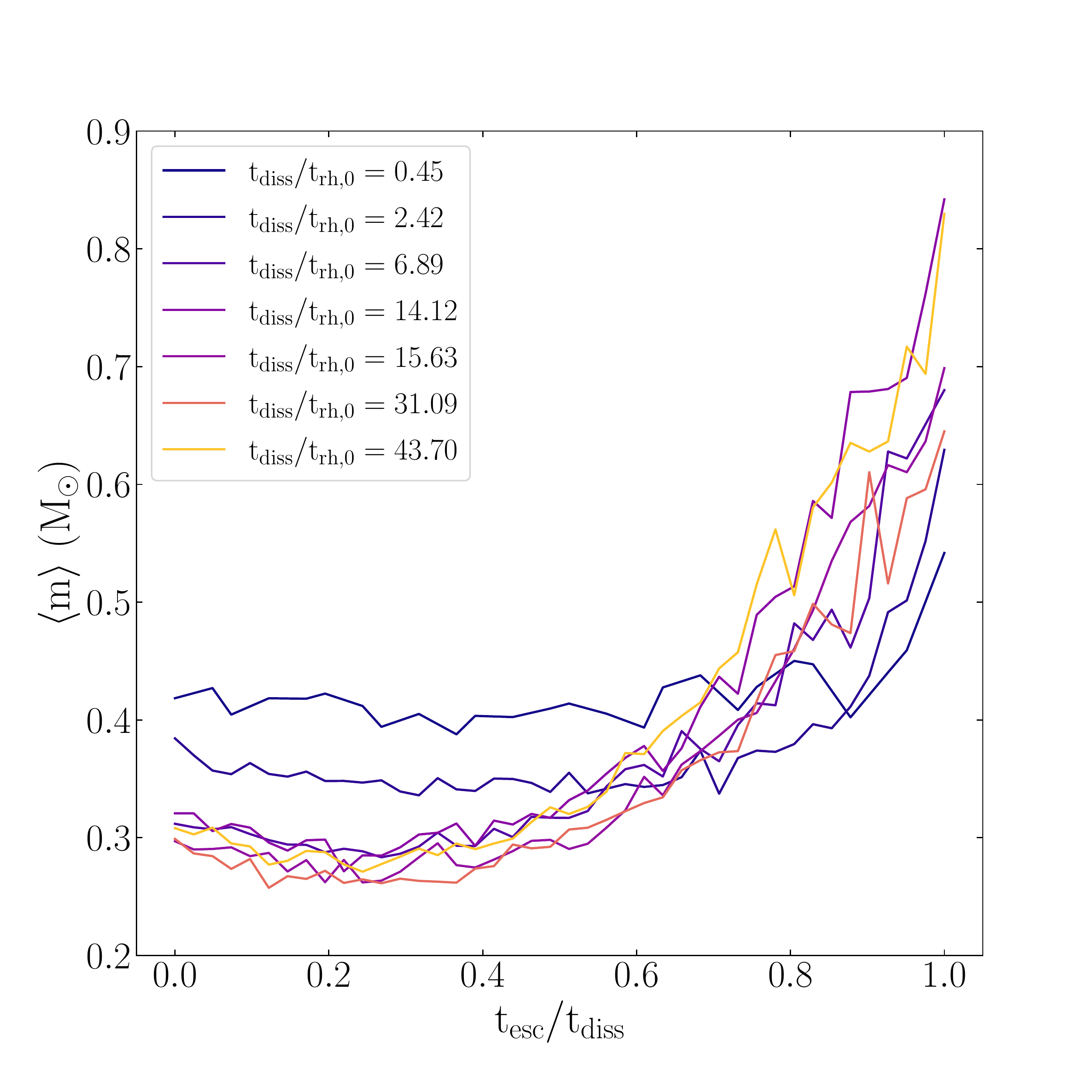}
    \caption{Mean mass $\langle m\rangle$ of escaping stars as a function of escape time normalized by cluster dissolution time for models initially consisting of 25,000 stars. Models are colour coded by the ratio of their dissolution time to their initial relaxation time, which is also given in the legend. Clusters with low $t_\mathrm{diss}/t_{rh,0}$ display little evolution in $\langle m\rangle$ with time, while clusters with high $t_\mathrm{diss}/t_{rh,0}$ show an initial decrease in  $\langle m\rangle$ followed by a steep increase as the cluster approaches dissolution.}
    \label{fig:mprof_tesc}
\end{figure}

We first consider how the mean mass of escaping stars changes as a function of time in each of our models in Figure \ref{fig:mprof_tesc}. Since each model cluster has a different dissolution time, we have normalized each star's escape time by the cluster's dissolution time for a better comparison. As mentioned above, only clusters that initially have 25,000 stars are shown here for clarity.

Figure \ref{fig:mprof_tesc} exemplifies how the properties of escaping stars change over time as a cluster evolves. For clusters with low $t_\mathrm{diss}/t_{rh,0}$, the mean mass of escaping stars stays relatively constant with time until it increases just before the cluster reaches dissolution, because the escaping stars are primarily high-mass stars that quickly segregated to the cluster's centre. Clusters with shorter dissolution times also have higher mean masses, because stars do not lose as much mass via stellar evolution before dissolution as they do in clusters with longer dissolution times.

\begin{figure*}
    \includegraphics[width=\textwidth]{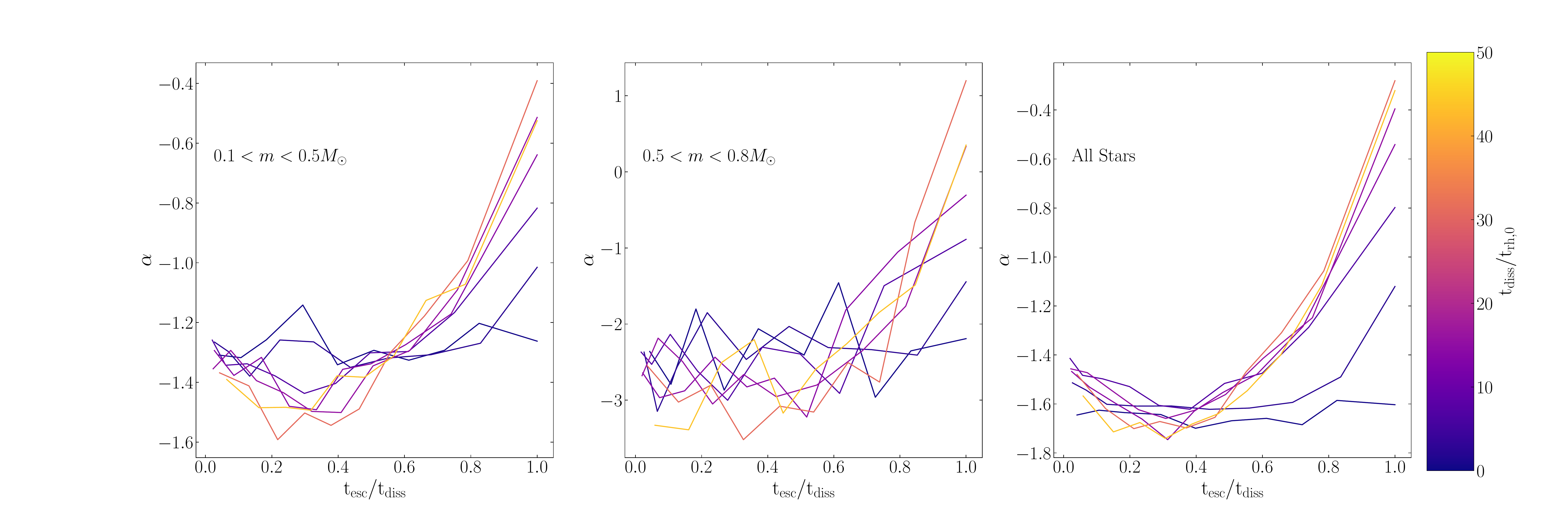}
    \caption{Slope of the stellar mass function of escaping stars $\alpha$ as a function of escape time normalized by cluster dissolution time for models initially consisting of 25,000 stars. The mass function is determined using stars between $0.1$ and $0.5\,M_{\odot}$ (left panel),  $0.5$ and $0.8\,M_{\odot}$ (middle panel), and all stars (right panel). Clusters with low $t_\mathrm{diss}/t_{rh,0}$ display little evolution in $\alpha$ with time, while clusters with high $t_\mathrm{diss}/t_{rh,0}$ show an initial decrease in  $\alpha$ followed by a steep increase as the cluster approaches dissolution. The trends have the same shape regardless of the mass range over which the slope $\alpha$ is measured.}
    \label{fig:aprof_tesc}
\end{figure*}

Clusters with high $t_\mathrm{diss}/t_{rh,0}$ are able to undergo significant relaxation before dissolution. In some cases, for clusters that are initially tidally under-filling, clusters can even undergo segregation before they start losing stars. Hence there is a small spread in the initial $\langle m \rangle>$ of model clusters with higher $t_\mathrm{diss}/t_{rh,0}$ ratios. Once stars begin escaping the cluster, the mean mass of escaping stars initially decreases as a function of time as low-mass stars continue to segregate outwards. Then, once clusters reach approximately $t_\mathrm{esc}/t_\mathrm{diss} = 0.4$, the mean mass of escaping stars begins to increase as the outer layers of the progenitor are stripped and high-mass stars which populated the inner regions of the progenitor begin escaping.

Figure \ref{fig:aprof_tesc}, which shows the evolution of the stellar MF of escaping stars' slope as a function of time, supports the above interpretation for the MF slope $\alpha$ measured across three different mass ranges. In each case, $\alpha$ remains constant in clusters with low $t_\mathrm{diss}/t_{rh,0}$ before increasing when the cluster approaches dissolution. Similarly $\alpha$ initially decreases for clusters with high  $t_\mathrm{diss}/t_{rh,0}$ before sharply increasing as the cluster approaches dissolution. There does not appear to be significant differences in the evolution of $\alpha$ for clusters with $t_\mathrm{diss}/t_{rh,0} > 20$. This result in consistent with \citet{Webb2016}, where it was found that radial variation in the stellar mass function within a cluster eventually reaches a maximum once clusters have experienced a significant number of relaxation times.

Comparing the different panels in Figure \ref{fig:aprof_tesc}, we see the expected trends are more clearly evident when considering low mass stars ($0.1$ to $0.5\,M_{\odot}$) or the entire mass spectrum. Intermediate mass stars between $0.5$ and $0.8\,M_{\odot}$ are both fewer in number (causing increased uncertainty in the calculation of $\alpha$) and segregate faster than low-mass stars. Hence there is very little difference in the stellar mass function of escaping stars for clusters with different $t_\mathrm{diss}/t_{rh,0}$ except when they are close to dissolution. With respect to observing stellar streams, we therefore only expect a variation in $\alpha$ to exist near where the progenitor would be located for stars between $0.5$ and $0.8\,M_{\odot}$. The rest of the stream will likely have little variation in $\alpha$.

\subsection{Properties of Stars as a Function of Distance Along the Stream} \label{ss_dprog}

\begin{figure}
    \includegraphics[width=0.48\textwidth]{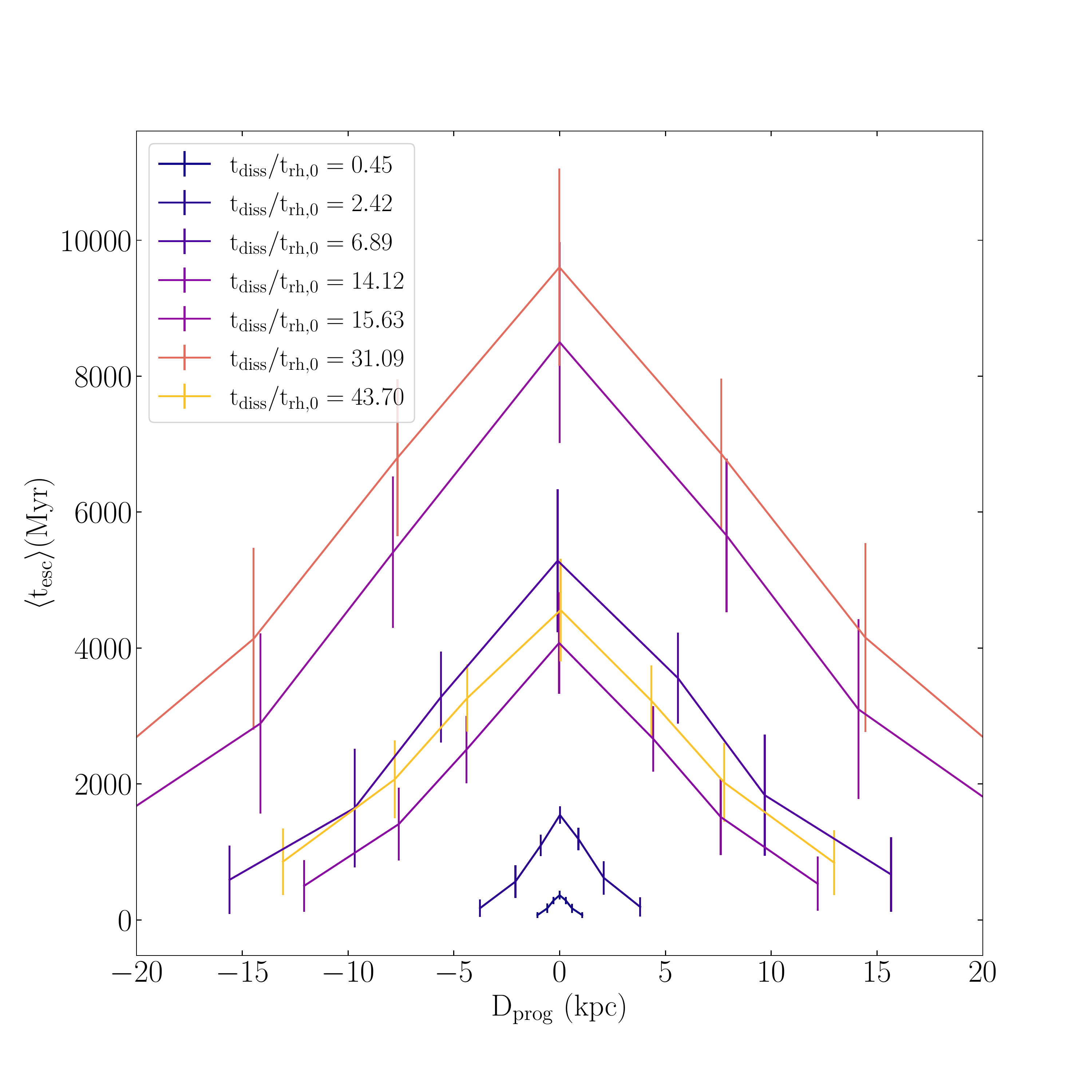}
    \caption{Mean escape time of stars in models initially consisting of 25,000 stars as a function of distance from the dissolved progenitor as measured along the stream path. Lines are colour coded by each cluster's $t_\mathrm{diss}/t_{rh,0}$ with each model's dissolution time marked in the legend. Error bars correspond to the error in the mean. The rate at which $t_\mathrm{esc}$ changes as a function of distance along the stream is, to first order, independent of progenitor dissolution time or $t_\mathrm{diss}/t_{rh,0}$. However, longer progenitor dissolution times lead to a large error in the mean due to stars mixing along the stream over time.}
    \label{fig:tesc_dprog}
\end{figure}

To determine whether the time evolution of the properties of escaping stars found in Section \ref{ss_tesc} translates to observational signatures for stellar streams, we first consider how well a star's position along a stream reflects the time at which it escaped the cluster. For each stellar stream, we find how the mean $t_\mathrm{esc}$ varies as a function of $D_\mathrm{prog}$ and calculate the error in the mean, with the results displayed in Figure \ref{fig:tesc_dprog}.

\begin{figure*}
    \includegraphics[width=\textwidth]{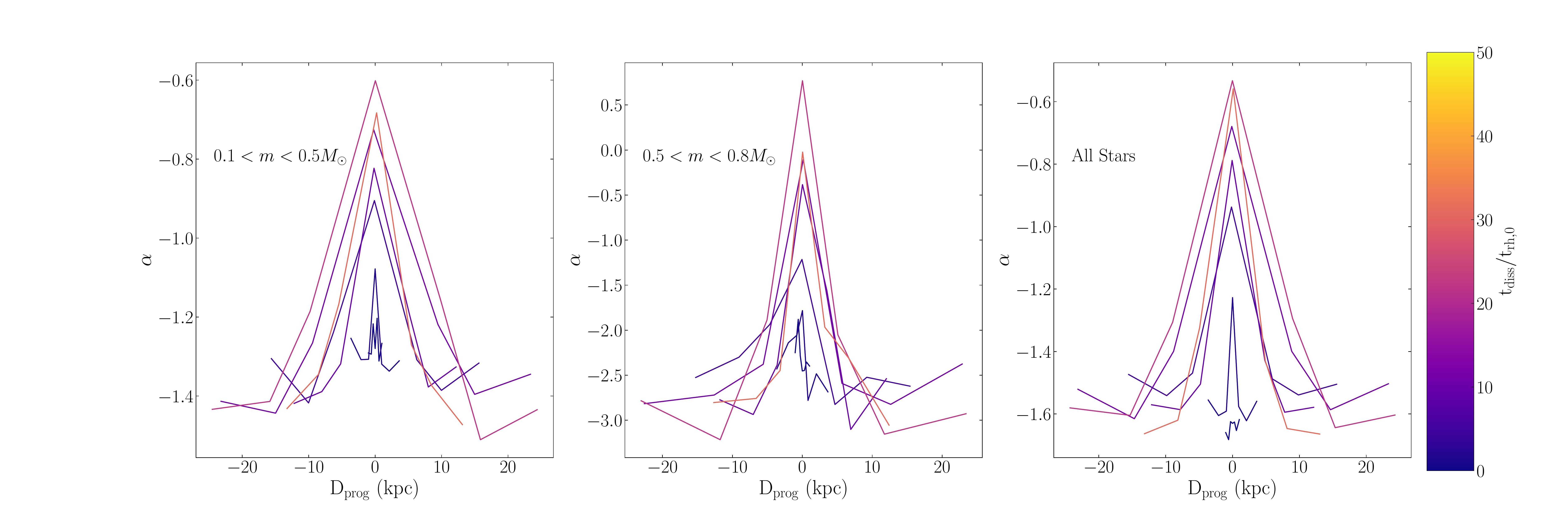}
    \caption{Slope $\alpha$ of the stellar mass function of escaping stars as a function of distance $D_\mathrm{prog}$ from the dissolved progenitor for models initially consisting of 25,000 stars. The mass function is determined using stars between $0.1$ and $0.5\,M_{\odot}$ (left panel),  $0.5$ and $0.8\,M_{\odot}$ (middle panel), and all stars (right panel). Clusters with low $t_\mathrm{diss}/t_{rh,0}$ display little evolution in $\alpha$ with $|D_\mathrm{prog}|$. Clusters with high $t_\mathrm{diss}/t_{rh,0}$ have a large maximum in $\alpha$ at the centre of the stream followed by a steep decrease as $|D_\mathrm{prog}|$ increases. Moving outwards, $\alpha$ eventually starts to increase again and then remain constant near the edges of the stream.}
    \label{fig:aprof_dprog}
\end{figure*}

Figure \ref{fig:tesc_dprog} illustrates that $D_\mathrm{prog}$ is an accurate proxy for the time at which a star escaped the progenitor cluster. The rate at which $t_\mathrm{esc}$ changes as a function of $D_\mathrm{prog}$ is, to first order, independent of both $t_\mathrm{diss}$ and $t_\mathrm{diss}/t_{rh,0}$ with a mean $d t_\mathrm{esc} / D_\mathrm{prog} \sim-300\,\mathrm{Myr\,kpc}^{-1}$. This result indicates that the mean escape velocity of stars is similar across all model clusters considered here. The dispersion in $t_\mathrm{esc}$ at a given $D_\mathrm{prog}$, however, does appear to depend on $t_\mathrm{diss}$. Progenitors with longer dissolution times appear to have larger dispersions in $t_\mathrm{esc}$, consistent with both the escape velocity dispersion increasing with cluster mass and stars along the stream having a longer time to mix as the progenitor dissolves. Hence, for clusters that dissolved over longer periods of times, $D_\mathrm{prog}$ is a slightly poorer proxy for $t_\mathrm{esc}$ than streams born out of clusters that dissolve quickly.

Having established that $D_\mathrm{prog}$ is an acceptable proxy for $t_\mathrm{esc}$, albeit with additional scatter in clusters with long dissolution times, we now explore how the MF slope $\alpha$ varies along streams as a function of $D_\mathrm{prog}$. Figure \ref{fig:aprof_dprog} demonstrates that the trends from Figure \ref{fig:aprof_tesc} between $\alpha$ and $t_\mathrm{esc}/t_\mathrm{diss}$ remain when considering $\alpha$ and $D_\mathrm{prog}$. Therefore, we conclude that stellar streams contain evidence for the dynamical state reached by its progenitor cluster before it dissolved. 

\section{Discussion}\label{s_discussion}

\begin{figure}
    \includegraphics[width=0.48\textwidth]{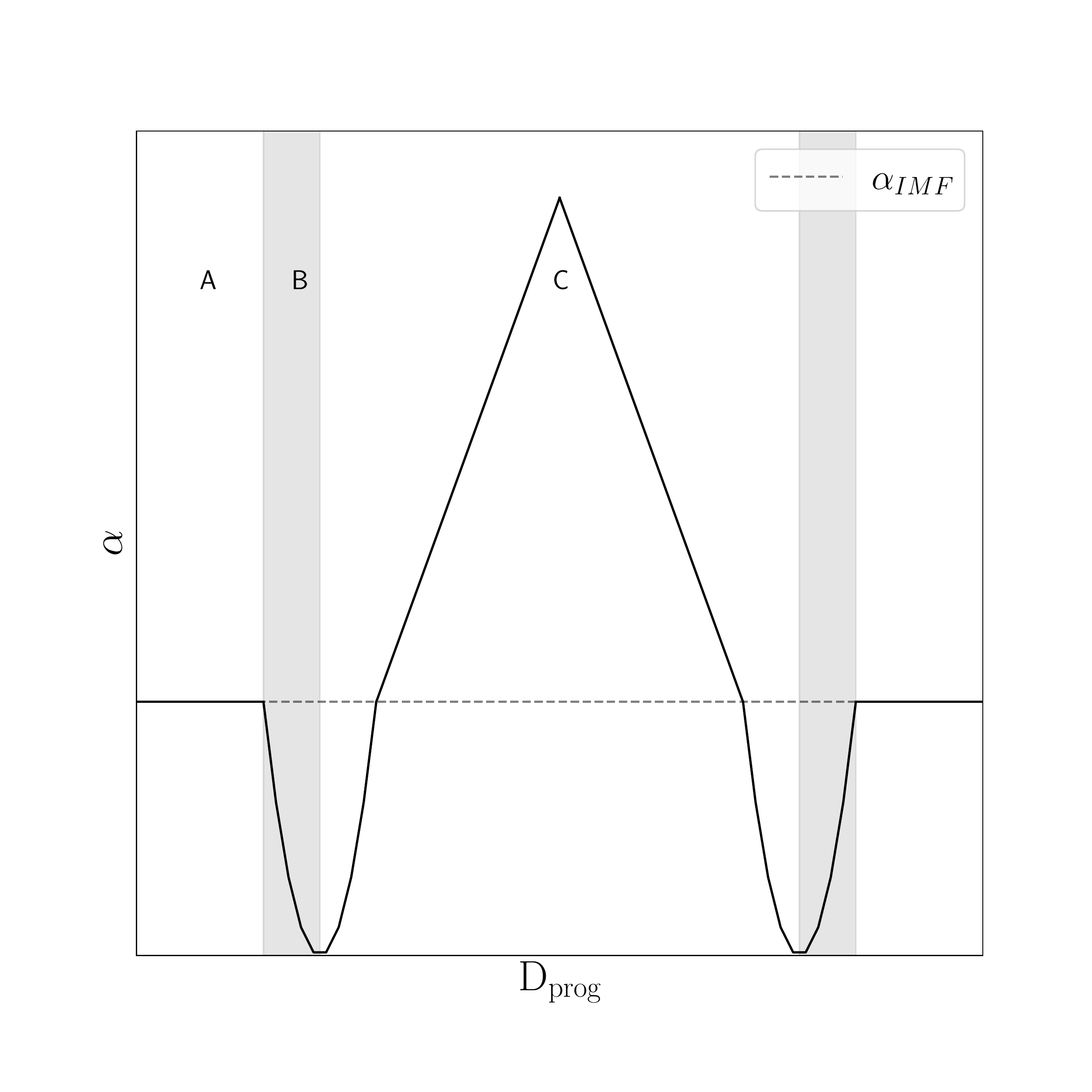}
    \caption{Schematic diagram illustrating how the slope $\alpha$ of the stellar mass function is expected to vary as a function of distance $D_\mathrm{prog}$ along a stellar stream formed through the dissolution of a dynamically-old star cluster. Stars in Region A are the first to escape the progenitor during an initial phase of unsegregated mass loss, such that the slope of the mass function in that region is equal to the primordial value $\alpha_{IMF}$. Stars in Region B escape the progenitor after a few relaxation times, while mass segregation is on-going and populating the outer regions with preferentially low-mass stars. Stars in Region C leave the progenitor as part of its dissolution after mass segregation has finished. The centre of the stream, where the progenitor cluster would be located had it not reached dissolution, corresponds to a peak in $\alpha$. }
    \label{fig:schematic}
\end{figure}

Through a suite of direct $N$-body simulations of star cluster evolution, we have established that information regarding the dynamical state reached by a given progenitor cluster is imprinted upon the stellar stream that emerges from the cluster's dissolution. Over the course of a cluster's lifetime, it goes through a stage of unsegregated mass loss if tidally filling, followed by segregated mass-loss and then complete dissolution. How these stages are reflected along a stellar stream is illustrated in Figure \ref{fig:schematic}, which features a schematic diagram of how the slope $\alpha$ of the stellar MF is expected to vary as a function of distance from the dissolved progenitor's location $D_\mathrm{prog}$. Assuming the cluster is initially tidally filling, the first stars to escape a cluster do so before any mass segregation has occurred, such that the mean mass and mass function of stars at the edges of a stream should reflect the cluster's initial mass function. For an under-filling cluster, if some segregation has occurred before dissolution begins, then $\alpha$ at the edges of the streams will be slightly lower than the primordial value. 

If a cluster is able to continue segregating before dissolution, the mean mass of escaping stars will decrease, as will the slope of the stellar mass function of escaping stars. Thus, at intermediate distances between a stream's edge and its centre a decrease in $\langle m\rangle$ and $\alpha$ will be observed. Finally, near the centre of the stream, the $\langle m\rangle$ and $\alpha$ increase sharply to a maximum as high-mass stars that quickly segregated to a cluster's core escape the cluster upon its dissolution. The above scenario implies that there should be a fundamental relationship between the range of mass function slopes reached along a stream relative to the dynamical state reached by the progenitor before dissolution. Furthermore, variation in the mass function along a stream should help in the identification of both the stream's edge and its centre. 

\subsection{Identifying the Edge of a Stream and the Location of the Progenitor's Remnant}

Our simulations have revealed that variation in the stellar mass function along a stellar stream should have a characteristic signature assuming the cluster had enough time to segregate before reaching dissolution. The signature, illustrated in Figure \ref{fig:schematic}, allows for key properties of the stream to be easily measured. For clusters that quickly become tidally filling after formation, such that stars began escaping the cluster before any segregation had occured, the mass function of the first stars to escape the cluster should reflect the IMF. Hence the MF of stars at the edge of a stellar stream should also be comparable to the IMF if the progenitor cluster was near tidally filling. The edges of the stream can therefore be determined to be where the MF is comparable to the IMF, $\alpha$ is constant or decreasing as distance from the centre of the stream decreases, and the stellar density is falling to zero. However if the progenitor cluster was tidally under-filling, such that stars didn't escape the cluster until after some segregation has occured, the slope of the mass function at the edges of the stream should be steeper (more negative) than the IMF. In the tidally under-filling case it is much harder to conclude the location of the stream edge. For example, it will not be possible to conclude whether or not the stream edge is where the density drops to zero and the MF is steeper than the IMF or if stars that would be located at the true stream edge have been perturbed or the stream density is too low. However in the case where the edges of the stream are steeper than the IMF while the overall mass function of the stream is consistent with the IMF, then it is likely the entire stream is being observed.

\begin{figure*}
    \includegraphics[width=\textwidth]{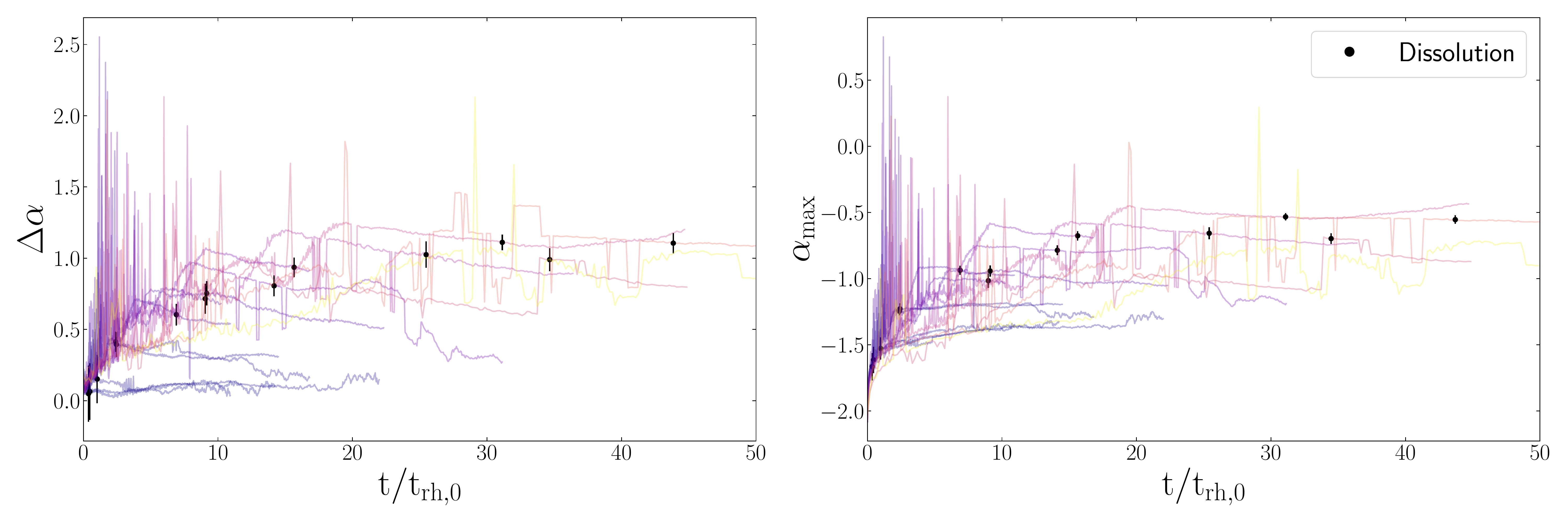}
    \caption{Left Panel: Difference between the maximum and minimum stellar mass function slope $\Delta \alpha$ measured along a stream as a function a time (normalized by initial relaxation time) for all stars in all models. The solid black points mark the value of $\Delta \alpha$ at dissolution, along with its uncertainty. Right Panel: Maximum stellar mass function slope $\alpha_\mathrm{max}$ measured along a stream as a function a time (normalized by initial relaxation time) for all stars in all models. While $\Delta \alpha$ appears to decrease with time after cluster dissolution, $\alpha_\mathrm{max}$ remains relatively constant to within measurement uncertainty until the entire stream disrupts. For comparison purposes, models are coloured based on $t_{diss}/t_{rh,0}$ as per Figure \ref{fig:aprof_dprog}.}
    \label{fig:alpha_time}
\end{figure*}

A significantly more rigorous constraint can be placed on the location of the stream's centre, as in almost all cases the centre of the stream where the progenitor's remnant is located corresponds to a sharp peak in the MF. This peak is due to the fact that high mass stars quickly segregated to the centre of the cluster and will be the last to escape as the cluster dissolves. The left panel of Figure \ref{fig:alpha_time} demonstrates the difference between the maximum and minimum stellar mass function slope $\Delta \alpha$ measured along a stream as a function a time for all stars in models with 25,000 stars. The figure demonstrates that the maximum slope remains separated from the minimum slope along the stream well after cluster dissolution. The difference does, however, decrease with time as stars mix along the stream.

In the right panel of Figure \ref{fig:alpha_time}, the time evolution of the maximum slope $\alpha_\mathrm{max}$ measured along the stream is illustrated. After dissolution, $\alpha_\mathrm{max}$ remains constant to within measurement uncertainty, which indicates  the minimum $\alpha$ increases with time as stars mix along the stream and causes $\Delta \alpha$ to increase. The fact that $\alpha_\mathrm{max}$ remains constant long after dissolution suggests the ability to identify the streams centre remains possible long after the progenitor as disrupted. We further observe that the $\alpha_\mathrm{max}$ reached by a given model cluster appears to be related to the number of initial relaxation times the cluster experiences before dissolution.

\subsection{Relating Variations in a Stream' Stellar Mass Function to its Progenitor}

In order to establish a relationship between variations in the stellar MF along a given stellar stream and the dynamical state reached by the stream's progenitor, we consider two different properties in our model stream. First, we explore how $\alpha_\mathrm{max}$ is related to the ratio between the each progenitor cluster's $t_\mathrm{diss}$ and initial half-mass relaxation time $t_{rh,0}$. Second, we measure the rate of change in $\alpha$ decreases as a function of $\ln D_\mathrm{prog}$ from the centre of the stream, \dalpha, out to either the minimum $\alpha$ reached along the stream or where $\alpha$ becomes constant with $D_\mathrm{prog}$.

\subsubsection{Relating $\alpha_\mathrm{max}$ to $t_\mathrm{diss}/t_{rh,0}$}

\begin{figure*}
    \includegraphics[width=\textwidth]{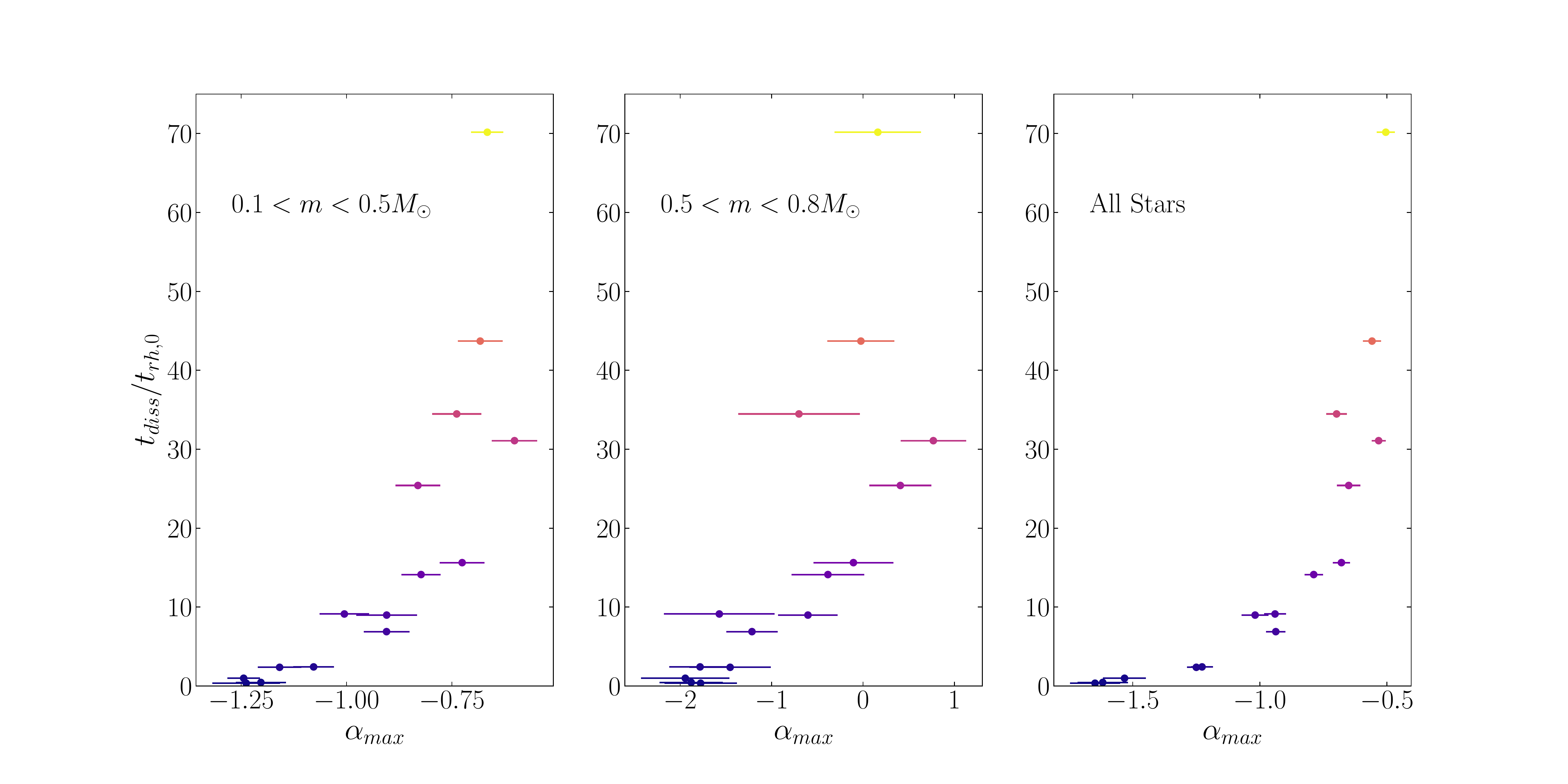}
    \caption{Progenitor cluster $t_\mathrm{diss}/t_{rh,0}$ as a function of the maximum stellar mass function slope $\alpha_\mathrm{max}$ at dissolution measured along each model stream for stars between 0.1 and 0.5 $M_{\odot}$ (left panel), 0.5 and 0.8 $M_{\odot}$ (middle panel) and all stars (right panel). Clusters with low $t_\mathrm{diss}/t_{rh,0}$ display a small increase in $\alpha$ at the centre of their resultant streams while clusters with large $t_\mathrm{diss}/t_{rh,0}$ show a large maximum $\alpha$ at the centre of their resultant streams. For $t_\mathrm{diss}/t_{rh,0} >$ 20, their appears to be no additional increase in maximum $\alpha$. For comparison purposes, models are coloured based on $t_{diss}/t_{rh,0}$ as per Figure \ref{fig:aprof_dprog}.}
    \label{fig:tdiss_tfunc_amax}
\end{figure*}

As discussed above, Figure \ref{fig:alpha_time} hints that $\alpha_\mathrm{max}$ provides an indication of the degree of mass segregation reached by a progenitor cluster before complete dissolution. Hence in Figure \ref{fig:tdiss_tfunc_amax} we consider $\alpha_\mathrm{max}$ at dissolution for each of our model streams as function of $t_\mathrm{diss}/t_{rh,0}$, where $\alpha_\mathrm{max}$ is measured over three different mass ranges as before. For each stream, stars are binned based on $D_\mathrm{prog}$ and separated into seven bins containing an equal number of stars.

Clusters with low $t_\mathrm{diss}/t_{rh,0}$ undergo very little relaxation before dissolution and therefore show small values of $\alpha_\mathrm{max}$. Conversely, clusters with large $t_\mathrm{diss}/t_{rh,0}$ underwent significant relaxation before dissolution. The rate at which $\alpha_\mathrm{max}$ increases with $t_\mathrm{diss}/t_{rh,0}$ is exponential, with $\alpha_\mathrm{max}$ varying very little for  $t_\mathrm{diss}/t_{rh,0} > 20$. This behavior is consistent with \citet{Webb2016}, where it was found that radial variation in a cluster's stellar mass function eventually stops increasing once clusters are sufficiently relaxed.

It is important to note, however, the exact values of $\alpha_\mathrm{max}$ in Figure \ref{fig:tdiss_tfunc_amax} strongly depend on how the data is binned. For example when keeping the number of stars in each bin the same, increasing the number of bins results in the central $\alpha$ increasing as only stars with later and later escape times remain. For model clusters with small values of $t_\mathrm{diss}/t_{rh,0}$ it is possible to converge upon a single value of $\alpha_\mathrm{max}$, but for dynamically evolved progenitor's no convergence is reached. A similar problem occurs when using bins of fixed width, which is then compounded by the fact that the value of $\alpha_\mathrm{max}$ continues to evolve past dissolution as the stream breaks up. Hence to properly constrain an observed stream's $t_\mathrm{diss}/t_{rh,0}$, a direct comparison between an observed dataset and simulations must be completed where simulated and observed data is binned the same way.

\subsubsection{Relating \dalpha\ to $t_\mathrm{diss}/t_{rh,0}$}

An alternative approach to using $\alpha_\mathrm{max}$ to constrain $t_\mathrm{diss}/t_{rh,0}$, that is less dependent on how the data is binned, is to use the rate of change in $\alpha$ as a function of $\ln D_\mathrm{prog}$. More specifically we take the absolute value of each star's $D_\mathrm{prog}$ and measure \dalpha\ between $\alpha_\mathrm{max}$ and either $\alpha_\mathrm{min}$ or to where $\alpha$ becomes constant as a function of $D_\mathrm{prog}$. Taking the absolute value of $D_\mathrm{prog}$ is acceptable since $\alpha(D_\mathrm{prog})$ is symmetric about $D_\mathrm{prog}=0$, with $\alpha (D_\mathrm{prog})$ decreasing linearly as function of $\ln D_\mathrm{prog}$.

\begin{figure*}
    \includegraphics[width=\textwidth]{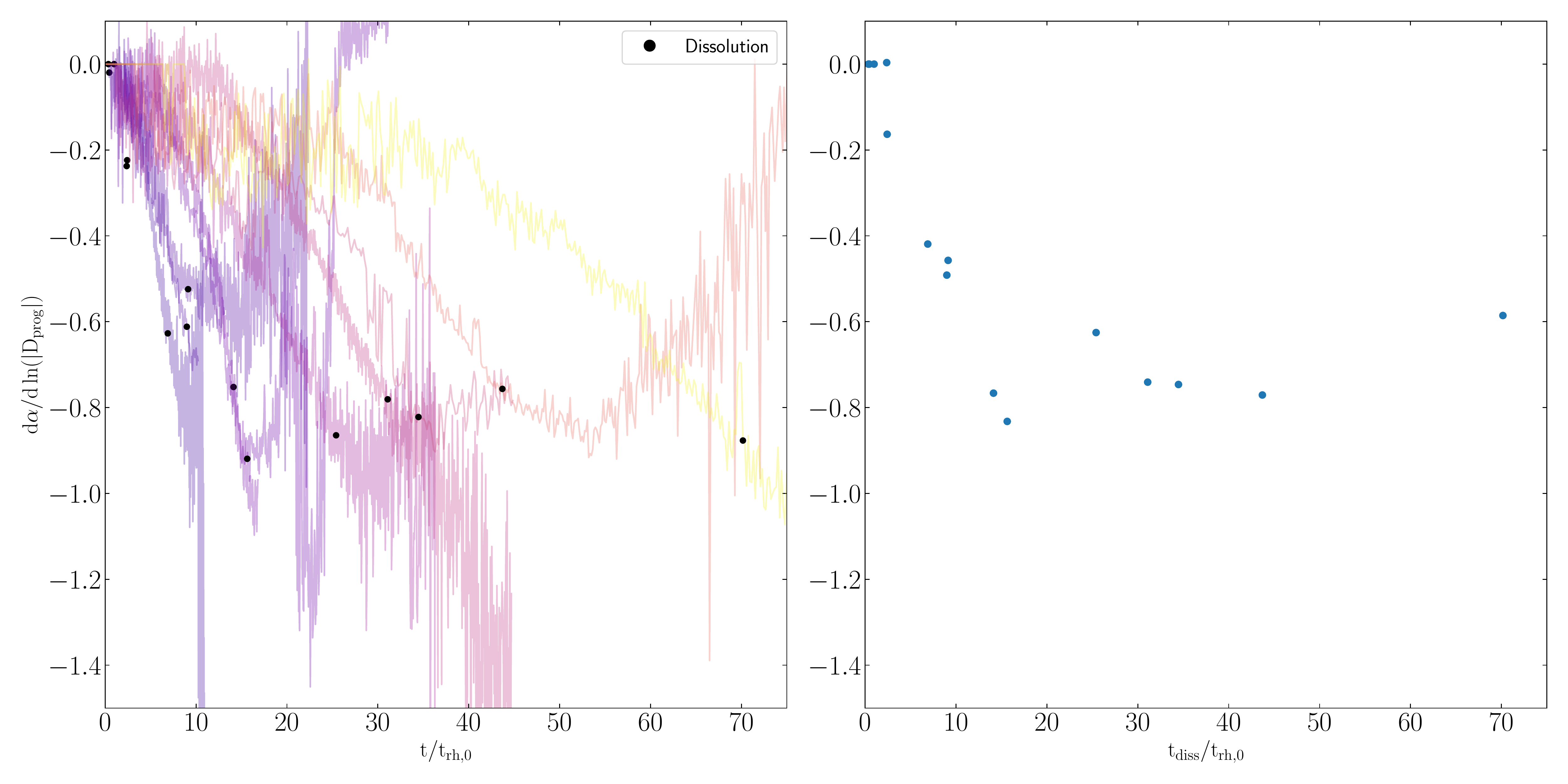}
    \caption{Left Panel: Time evolution of the rate of change in $\alpha$ with respect to $\ln D_\mathrm{prog}$ between the point along the stream where the stellar mass function slope reaches a maximum and the point along the stream where the slope reaches are minimum or begins to remain constant with $\ln D_\mathrm{prog}$. The solid black points mark the value of \dalpha\ at dissolution, with uncertainties smaller than the datapoints. It should be noted that time is normalized by each cluster's and the mass function is measured using all stars. While the rate of change at dissolution in shown for all models, the time evolution of models with $t_\mathrm{diss}/t_{rhh,0} < 2.5$ are not shown due to the significant fluctuation in \dalpha. For comparison purposes, models are coloured based on $t_{diss}/t_{rh,0}$ as per Figure \ref{fig:aprof_dprog}. Right Panel: \dalpha\ at dissolution time as a function of $t_\mathrm{diss}/t_{rhh,0}$ for each model cluster. In all cases, \dalpha\ decreases as a function of time, reaching more negative values for progenitors that experience several relaxation times before dissolution.}
    \label{fig:dalpha_time}
\end{figure*}

The left panel of Figure \ref{fig:dalpha_time} illustrates the time evolution of \dalpha\ for all models with $t_\mathrm{diss}/t_{rhh,0} > 2.5$, with time normalized by $t_{rh,0}$ and $\alpha$ measured using all stars in the cluster. For a homogenous measurement between models, stream stars are binned by $|D_\mathrm{prog}|$ in bins of width 0.5 kpc. The value of \dalpha\ at dissolution is marked for all models as a solid black point, with the right panel of Figure \ref{fig:dalpha_time} illustrating each cluster's \dalpha\ at dissolution as a function of $t_\mathrm{diss}/t_{rhh,0}$. At early times there is significant fluctuation in \dalpha\ as the tidal tails grow, since \dalpha\ is only calculated using a small number of bins. The time evolution of models with $t_\mathrm{diss}/t_{rhh,0} < 2.5$ are not illustrated in the left panel as \dalpha\ fluctuates even more significantly between timesteps since the streams are so short and there is little actual variation in $\alpha$ along the stream.

Figure \ref{fig:dalpha_time} demonstrates how the rate of change in $\alpha$ along the stream gets steeper with time for all clusters. For clusters that dissolve quickly,such that $t_\mathrm{diss}/t_{rhh,0}$ is small, the variation in $\alpha$ along the stream in minimal. Clusters that dissolve after several relaxation times, on the other hand, can develop a strong gradient in $\alpha$ along the stream. In fact, focusing on the value of \dalpha\ at each cluster's dissolution time in the right panel of Figure \ref{fig:dalpha_time},  \dalpha\ at dissolution decrease exponentially as a function of $t_\mathrm{diss}/t_{rhh,0}$.

Beyond dissolution, the left panel of Figure \ref{fig:dalpha_time} further shows that \dalpha\ will continue to evolve. This evolution is due to stars along the stream continuing to move away from the stream's centre, where the progenitor would be located. Hence each star's $D_\mathrm{prog}$ increases over time as the stream breaks apart. $\alpha_{max}$ will also increase in the centre over time, as only stars that orbited in the clusters centre will remain in the central bin, but at a slower rate than $D_\mathrm{prog}$ increases. The fact that \dalpha\ continues to evolve after dissolution means that a direct calculation of a progenitor's $t_\mathrm{diss}/t_{rhh,0}$ can't be made from a measurement of \dalpha\ alone. Instead, observational datasets must be compared directly to models that have been put through the same analysis.

\subsection{Application to Globular Clusters with Tidal Tails}\label{s_pal5}

In order to explore the applicability of our results to globular clusters with observed tidal tails, we further apply our analysis to an $N$-body simulation of the Milky-Way globular cluster Pal 5 and its tidal tails \citep{Odenkirchen2001, 2002AJ....124..349R}. The mock Pal 5 stream is generated using the same direct $N$-body code as our suite of simulations, \texttt{NBODY6} \citep{aarseth03}; this simulation was originally presented in \citet{Starkman2020}. The simulation consists of initially 100,000 stars in the form of a Plummer model \citep{Plummer1911} with a half-mass radius of 10 pc. All other features of the simulation are identical to those presented in Section \ref{s_nbody}. The cluster's initial position in the Galaxy was determined by reverse integrating Pal 5's current position and velocity from \citet{Vasiliev2019} by 12 Gyr in the \texttt{MWPotential2014} gravitational potential. After 12 Gyr, the simulated cluster's mass and stellar MF are comparable to the observed properties of Pal 5 \citep{grillmair16, Ibata2016, Ibata17, erkal17, Bonaca2020}. The cluster is, however, slightly more compact than the latest observations of Pal 5. Recently, \citet{Gieles2021} also produced a Pal5-like star cluster using direct $N$-body simulations with a star cluster that is initially more compact than our simulations and closer to the current size of Pal 5 at the end of the simulation. However, in our simulations black holes are kicked from the cluster with a high velocities while \citet{Gieles2021} retains black holes and show they play a pivotal role in Pal 5's evolution. The impact of black hole retention on measuring mass segregation along the stream, however, is negligible.

\begin{figure}
    \includegraphics[width=0.48\textwidth]{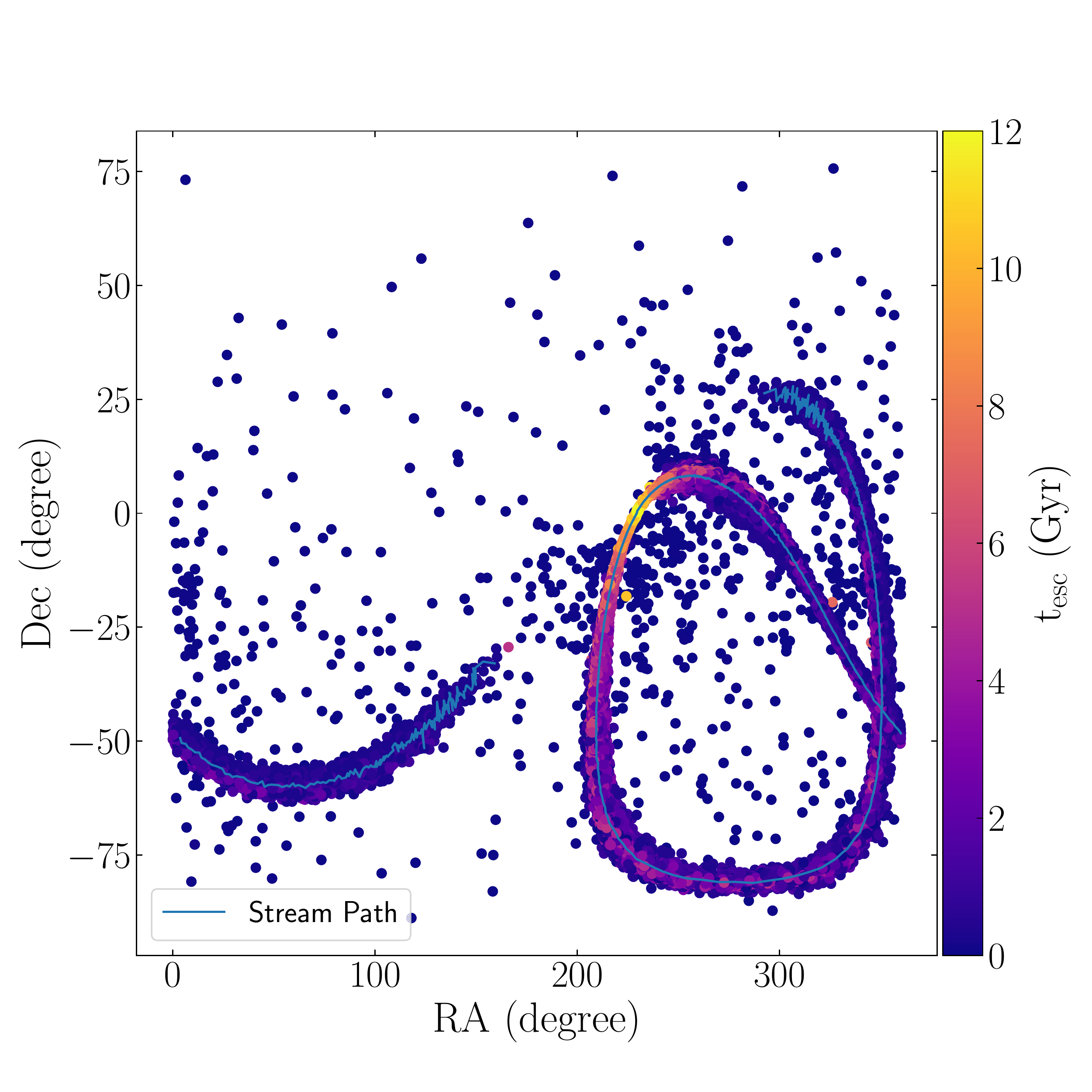}
    \caption{RA and Dec of tail and cluster stars in a Pal5-like $N$-body model. Stars are colour coded based on the time at which they escape the cluster, as indicated by the colour bar and the stream path is shown in blue.}
    \label{fig:pal5_obs}
\end{figure}

Figure \ref{fig:pal5_obs} illustrates the on-sky coordinates of model stars that make up the tidal tails and the cluster itself, colour coded by each star's escape time. Given that Pal 5 has an orbital eccentricity of $\sim 0.14$ and that the stream path is projected onto the plane of the sky, there can be significant contamination along the stream path where the orbital path crosses itself. Stream stars that escaped the cluster long ago can have projected locations that are similar to stars that escaped the cluster recently, while their true three-dimensional separation is much larger. Hence measuring how any property varies along the stream path will have considerably more uncertainty than in the idealised test cases presented in Section \ref{s_results}, as $D_\mathrm{prog}$ will not always directly correlate with $t_{esc}$.

It is also important to note that the simulated tails are significantly longer than the observed Pal 5 system, believed only to extend $\pm 15$ degrees from the cluster \citep{Starkman2020}. Since the model cluster spends its entire lifetime in the Milky Way and there is no contamination from field stars, tail stars can be observed over 360 degrees away from the cluster as measured along the tail path. Since the purpose of this analysis is to simply explore how the mass function varies along stellar streams with realistic orbits, we address the possibility of actually measuring a variation about Pal 5 in Section \ref{s_obs}.

\begin{figure*}
    \includegraphics[width=\textwidth]{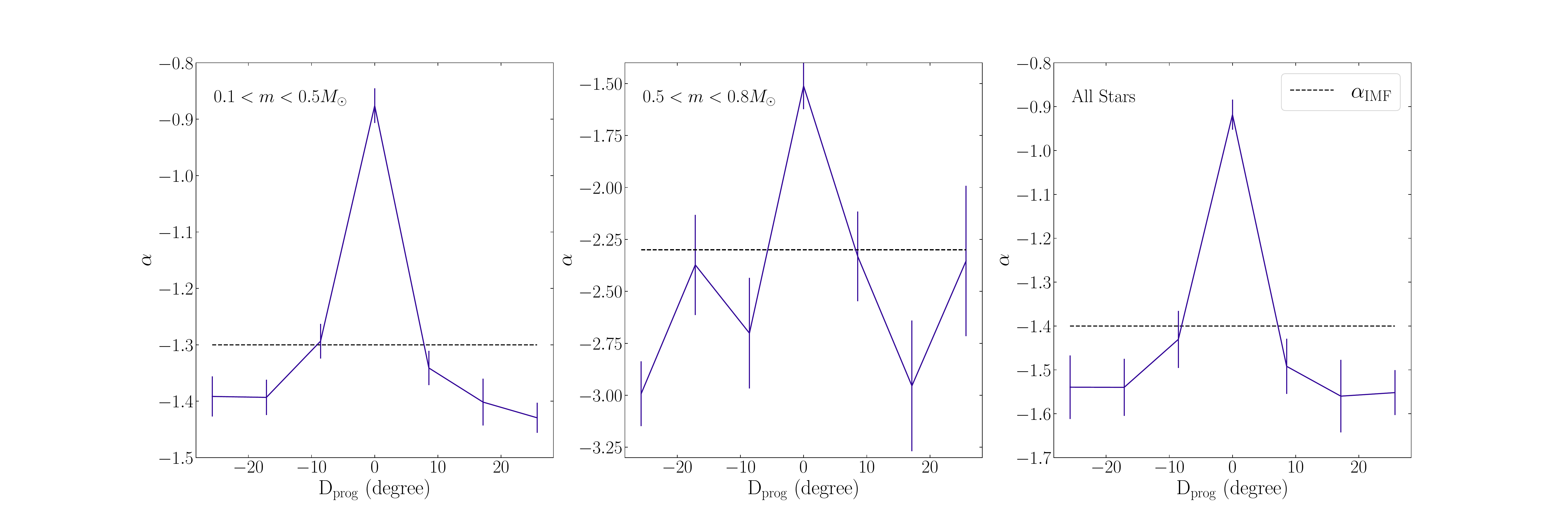}
    \caption{Slope $\alpha$ of the stellar mass function of escaping stars as a function of distance from the centre of the stream for a model of the Pal 5 cluster and its tidal tails. The mass function is determined using stars between 0.1 and 0.5 $M_{\odot}$ (left panel),  0.5 and 0.8 $M_{\odot}$ (middle panel), and all stars (right panel). The initial value of $\alpha$ in each mass range is marked with a horizontal dotted line. In all three panels the general relationship between $\alpha$ and $D_\mathrm{prog}$ found in the suite of simulations is reproduced.}

    \label{fig:pal5_obs_error}
\end{figure*}

Figure \ref{fig:pal5_obs_error} illustrates our measurements of $\alpha$ along our simulated Pal 5 stream, as a function $D_\mathrm{prog}$ as measured on the plane of the sky, for three different mass ranges. For all three mass ranges, we recover the expected trend of how $\alpha$ varies along a stream produced by a dynamically evolved progenitor cluster. Hence both the centre of the stream and its edges can be recovered for progenitor cluster's on non-circular orbits and when restricted to on-sky stellar positions. This statement holds even when the simulation is evolved further to the point where Pal 5 becomes completely unbound. 

Measurements of $\alpha_\mathrm{max}$ and \dalpha\ for the Pal 5 system are also consistent with Figures \ref{fig:tdiss_tfunc_amax} and \ref{fig:dalpha_time}. Considering the right panel of Figure \ref{fig:pal5_obs_error} where $\alpha$ is measured using all stars, Pal 5 has an $\alpha_\mathrm{max}$ of approximately $-0.9$. This maximum slope is consistent with models at dissolution in Figure \ref{fig:tdiss_tfunc_amax}, with the Pal 5 model having $t_\mathrm{diss}/t_{rhh,0} = 4.6$ when data is binned the same as it was for our suite of simulations (7 bins containing an equal number of stars, centred on $\rm D_\mathrm{prog} = 0$). Similarly we measure $\rm d \alpha / d \ln (|D_\mathrm{prog}|)=-0.15$ for the Pal 5 model, which is consistent with models at dissolution in Figure \ref{fig:dalpha_time} and the Pal 5 model having $t_\mathrm{diss}/t_{rhh,0} = 4.6$, when the tails are separated into $|D_\mathrm{prog}|$ bins of width 0.5 kpc. The agreement between our Pal 5 model and the relationships observed in Figures \ref{fig:tdiss_tfunc_amax} and \ref{fig:dalpha_time} for model clusters at dissolution is likely due to the fact that Pal 5 is approaching dissolution itself \citep{Gieles2021} and its current relaxation time is so large ($10^{9.87}$ years) that it will undergo little dynamical evolution before dissolving \citep{Baumgardt2017}.

\subsection{Application to Observations}\label{s_obs}

\begin{figure*}
    \includegraphics[width=\textwidth]{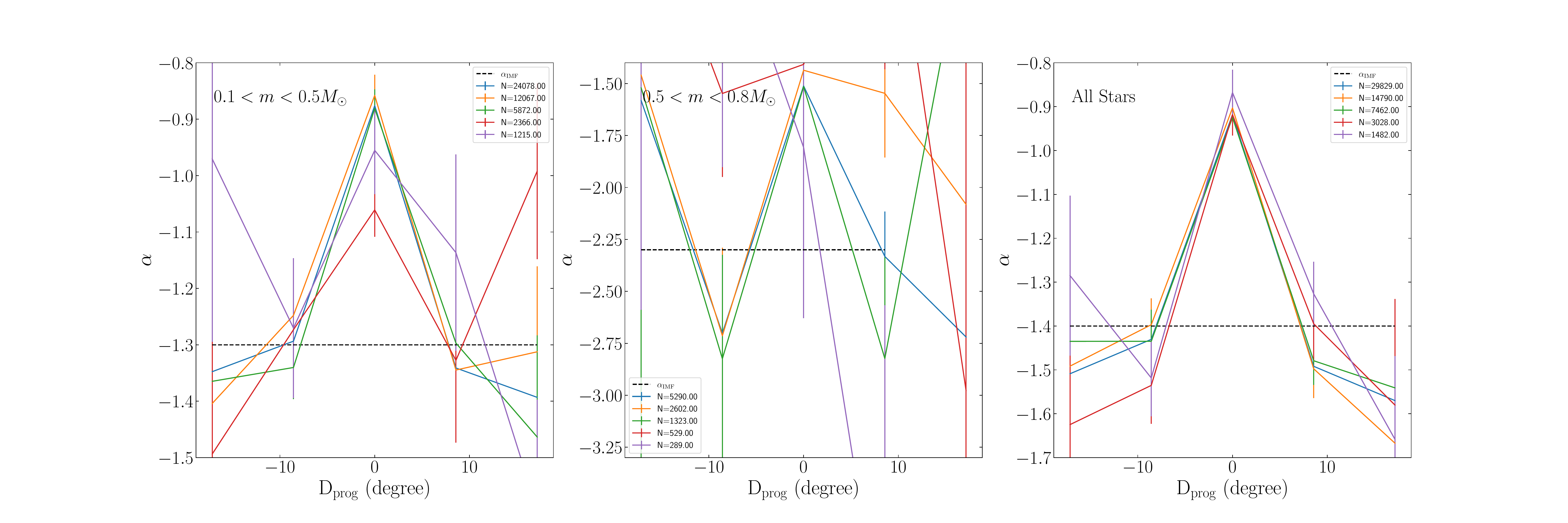}
    \caption{Slope $\alpha$ of the stellar mass function of escaping stars as a function of distance from the centre of the stream for a model of the Pal 5 cluster and its tidal tails for stars within 15 degrees of Pal 5. The mass function is determined using stars between 0.1 and 0.5 $M_{\odot}$ (left panel),  0.5 and 0.8 $M_{\odot}$ (middle panel), and all stars (right panel) assuming $100\%$ (blue), $50\%$ (orange),$25\%$ (green),$10\%$ (red) and $5\%$ (purple) completeness. The exact number of stars used to measure the profile in each case are indicated in the legend. Decreasing the completeness minimizes the ability to observe variation in the mass function along the stream.}

    \label{fig:pal5_obs_sampling}
\end{figure*}

The previous analysis of our Pal 5-like model cluster assumed that tail stars are recovered with $100\%$ completeness. In reality, the tail stars of Pal 5 are only observed within 15 degrees of the cluster \citep{Starkman2020} with unknown completeness. To explore the effects of a limited stream length and smaller tail star population, we re-examine the Pal 5 model using only stars within 15 degrees of the cluster and sub-sample the population assuming  $100\%$, $50\%$,$25\%$,$10\%$ and $5\%$ completeness. The variation in $\alpha$ along the stream, given these limitations, is illustrated in Figure \ref{fig:pal5_obs_sampling}. Note that the y-axis limits of Figure \ref{fig:pal5_obs_sampling} have been set to match Figure \ref{fig:pal5_obs_error}.

In each panel of Figure \ref{fig:pal5_obs_sampling}, the uncertainty in each measurement of $\alpha$ expectantly increases as the number of stars used to calculate $\alpha$ decreases. Hence the ability to confidently measure variation in $\alpha$ along the tails decreases as well. When only using stars between 0.1 and 0.5 $M_{\odot}$ (left panel), a similar $\alpha$ profile is found for completeness levels between 50 and 100 percent (at least 5991 stars). However below $50\%$, only the peak $\alpha$ is comparable to the higher completeness cases. When only using stars between 0.5 and 0.8 $M_{\odot}$ (middle panel), the $\alpha$ profile is only recovered for completeness levels above $75\%$ (at least 2753 stars). It is only when using all stars in the cluster that the $\alpha$ profile is recovered down to 5 percent completeness (1494 stars). Hence, even if variation in $\alpha$ exists along a stellar stream, several thousand stars are potentially required to observe the variation, depending on the dataset's limiting magnitude.

\section{Conclusion}\label{s_conclusion}

By analyzing model stellar streams born out of progenitor clusters that formed with a range of half-mass relaxation times and dissolution times, we have shown that the dynamical state reached by the progenitor cluster before dissolution is imprinted upon the resultant stellar stream. The signature lies in how the stellar mass function varies along the stream, between the stream's edges and its centre, and depends on how many relaxation times a cluster can experience before reaching dissolution.

The distribution of stellar masses at the leading and trailing edges of the stream should reflect the cluster's initial MF, regardless of the dynamical state reached by the progenitor, as these are the first stars to escape the cluster before it has undergone any mass segregation. If the progenitor cluster dissolved quickly relative to its initial relaxation time, such that very little mass segregation occurred, then the distribution of stellar masses observed at the edges of the stream will be the same across the entirety of the stream. If, however, the progenitor cluster survived to evolve for several relaxation times before dissolution, the mean mass of escaping stars will have at some point decreased. Therefore, moving inwards from the edges of the stream, it is expected that the slope of the stellar mass function will decrease to reflect the fact that preferentially low-mass stars were at one point escaping the cluster. The minimum $\alpha$ reached along the stream, and the distance at which that minimum is located with respect to the edges of the stream, will depend on the dynamical age reached by the progenitor before dissolution. If the progenitor remained dynamically young, only a small decrease in $\alpha$ will be observed over a small fraction of the stream. If the progenitor evolved to the point of being dynamically old, a larger decrease in $\alpha$ is expected over a larger fraction of the stream.

After reaching a minimum, $\alpha$ is expected to rise again when moving inwards along the stream, reaching a peak at the centre of the stream where the progenitor cluster would be located had it not dissolved. The peak is a direct result of high-mass stars segregating to the progenitor's centre at a much faster rate than low-mass stars segregate outwards. The rate at which $\alpha$ increases with respect to location along the stream will again depend on the dynamical state reached by the progenitor. If the progenitor dissolved while still dynamically young, perhaps only high-mass stars will have segregated to the cluster's centre, resulting in a gradual increase in $\alpha$. However if the progenitor becomes so dynamically old that segregation stops, as found by \citet{Webb2016}, the mean mass of escaping stars will have continued to increase as the cluster approached dissolution. A continual increase in the mean mass of escaping stars will be reflected by a steep increase in $\alpha$ along the stream towards the peak at the centre.

The characteristic signature outlined above, and previously illustrated in Figure \ref{fig:schematic} means that we can use a measurement of the radial variation in the stellar mass function along a stream to constrain properties of the stream and its progenitor. With respect to the stream itself, we can determine whether or not the entire stream is being observed if the mass function of stars at the edges of the stream reflect the expected IMF (allowing for the effects of stellar evolution) and the slope of the mass function is either constant or decreasing as distance from the centre of the stream decreases. Hence for streams like GD-1, where it has yet to be confirmed that the edges of the stream have been observed \citep{deboer18, PriceWhelan2018}, measurements of how the mass function varies along the stream will be extremely useful. Furthermore a sharp increase in the slope of mass function $\alpha$, which indicates that the region is dominated by higher mass stars, corresponds to the location of the stream's centre. Knowing a stream's centre and its edges are essential in accurately modelling stellar streams \citep[e.g.,][]{Varghese2011, Bovy14a, Bonaca2014, PriceWhelan2014, sanders16, bonaca18, Webb2019}.

With respect to the progenitor cluster itself, we find that the slope of the mass function at the peak ($\alpha_\mathrm{max}$) is related to the number of initial relaxation times a cluster experiences before dissolution. However, the measured relationship between $\alpha_\mathrm{max}$ and $t_\mathrm{diss}/t_{rhh,0}$ depends strongly on how the data is analysed. Alternatively, the rate at which $\alpha$ decreases along the stream from the peak to either its minimum value or to where $\alpha$ remains constant with distance from the progenitor (\dalpha) is also related to $t_\mathrm{diss}/t_{rhh,0}$. When measured at dissolution, \dalpha\ decreases exponentially as $t_\mathrm{diss}/t_{rhh,0}$. While the measured relationship between \dalpha\ and $t_\mathrm{diss}/t_{rhh,0}$ at cluster dissolution is independent of how the data is analysed, \dalpha\ will continue to evolve after dissolution which makes it difficult to use as a proxy for $t_\mathrm{diss}/t_{rhh,0}$ for any given stream. Hence direct comparisons between models and observations are required in order to constrain a given stream's $t_\mathrm{diss}/t_{rhh,0}$, unless it can be determine that the stream just recently dissolved. In our application to a Pal 5-like model cluster, a direct comparison between the Pal 5 model and our suite of simulations results in both $\alpha_\mathrm{max}$ and \dalpha\ correlating with the the cluster's $t_\mathrm{diss}/t_{rhh,0}$ since it is quite close to dissolution and will undergo little evolution before breaking up.



Given the properties of clusters in our suite of simulations, we can only conclude that the results presented here are applicable to cluster's with $t_{rh,0} \lesssim 1\,\mathrm{Gyr}$ and $t_\mathrm{diss} \lesssim 12\,\mathrm{Gyr}$. Taking the approximation for cluster dissolution time in \citet{Baumgardt03} and the globular cluster mass-size relationship presented in \citet{Choksi2019}, we can expect similar results for streams orbiting within 10 kpc of the Galaxy's centre with progenitors that had initial masses less than $2. \times 10^4 M_{\odot}$. For streams orbiting near 20 kpc, comparable to the orbits of Pal 5 and GD-1, the results will apply if the streams progenitor had an initial mass less than $7 \times 10^3 M_{\odot}$. However, the fact that similar behaviour was observed in our Pal 5 model, which has an initial relaxation time of 2.7 Gyr and dissolution time beyond 12 Gyr suggests our results may also apply to streams with larger and more massive progenitors. 

It is important to note that the model clusters explored in this study, with the exception of Pal 5, all have circular orbits. While the Pal 5 model illustrated that parts of a given stream can overlap with itself when the stream has an eccentric orbit, the length of the stream will also vary with orbital phase \citep{kaderali18}. Hence it must be determined how $\alpha_\mathrm{max}$ and \dalpha\ vary with orbital phase for a given orbit in order to directly apply to the results to a given stream. Furthermore, how the results of our study apply to streams that experience strong perturbation due to substructure \citep[e.g.,][] {Ibata2001, johnston02, erkal15a, erkal15b, Amorisco16a, bovy17, pearson17, banik18, Bonaca2019} and streams with progenitor clusters that were accreted by the Milky Way \citep{Carlberg2018, Malhan2019, Carlberg2020, Bonaca2021} must also be considered.

The ability to measure the stellar mass function along a stellar stream is currently limited to only higher-mass stars when using data from large-scale surveys. The \emph{Gaia} satellite \citep{gaia18}, which is only complete between G=12 and G=17, will only recover stars down to a minimum mass of $\sim 0.7 M_{\odot}$ assuming an isochrone \citep{Bressan12a} and distances corresponding to GD-1 \citep{Webb2019} and Pal 5 \citep{Vasiliev2019}. The Sloan Digital Sky Survey is capable of reaching masses between $\sim 0.5$ and $0.7 M_{\odot}$ for GD-1 and Pal 5 like distances \citep{Bilir2009, Blanton2003}, but it is much harder to confirm stream membership without kinematic information as well. Dedicated Hubble Space Telescope time could resolve stars down to the masses required to accurately measure variation in the mass function along a stream, but again it would be difficult to confirm stream membership. Upcoming surveys with the next generation of telescopes, like the Vera C. Rubin Observatory \citep{verarubin}, Euclid \footnote{https://www.euclid-ec.org/}, or the European Extremely Large Telescope \footnote{https://www.eso.org/sci/facilities/eelt/} will also likely allow for radial variation in the stellar mass function to be measured along streams. Alternative tracers related to the progenitor's dynamical age, like the distribution of blue straggler stars \citep{Ferraro2012} and the degree of multiple population mixing \citep{Vesperini2013}, may also be reflected in the distribution of stars along a stream.

The analysis performed in this study provides theoretical predictions for how we expect the stellar mass function to vary along stellar streams that the upcoming telescopes mentioned above will be able to confirm. From this work, it will be possible to better model observed stellar streams and place constraints on the progenitor clusters that dissolved into the stellar streams that we observe in the Galaxy today. Revealing the Milky Way's dissolved star cluster population is an important step towards being able to rewind our own Galaxy's evolution, its merger history, and understanding star cluster formation.

\section*{Acknowledgements}

The authors would like to thank the anonymous referee for helpful feedback on our manuscript. JB acknowledges financial support from NSERC (funding reference number RGPIN-2020-04712) and an Ontario Early Researcher Award (ER16-12-061). This work was made possible in part by the facilities of the Shared Hierarchical Academic Research Computing Network (SHARCNET: \url{www.sharcnet.ca}) and Compute/Calcul Canada.

\section*{Data Availability}
The simulated data underlying this article will be shared on reasonable request to the corresponding author.

\bibliographystyle{mnras}
\bibliography{ref2}

\bsp

\label{lastpage}

\end{document}